\documentclass[aps,prl,twocolumn,superscriptaddress]{revtex4}

\usepackage{graphicx,psfrag}
\usepackage[latin1]{inputenc}
\usepackage{natbib}
\usepackage{amsmath, amsthm, amssymb}

\newcommand{\ds}{\displaystyle}
\newcommand{\mb}{\mathbf}
\newcommand{\notes}[1]{}

\newcommand{\beq}{\begin{equation}}
\newcommand{\eeq}{\end{equation}}
\newcommand{\beqnn}{\begin{equation*}}
\newcommand{\eeqnn}{\end{equation*}}
\newcommand{\beqas}{\begin{eqnarray*}}
\newcommand{\eeqas}{\end{eqnarray*}}
\newcommand{\beqa}{\begin{eqnarray}}
\newcommand{\eeqa}{\end{eqnarray}}
\newcommand{\p}{\partial}

\begin{document}

\title{Spin wave excitation patterns generated by spin torque oscillators}
\author{F. Maci\`a$^*$}
\address{Department of Physics, New York University, 4 Washington Place, New York, NY 10003}
\author{F.C. Hoppensteadt}
\address{Courant Institute of Mathematical Sciences, New York University, 251 Mercer Street, New York, NY 10012}
\author{A.D. Kent}
\address{Department of Physics, New York University, 4 Washington Place, New York, NY 10003}
\date{\today}

\begin{abstract}
Spin torque nano oscillators (STNO) are nano-scale devices that can convert a direct current into short wave-length spin-wave excitations in a ferromagnetic layer. We show that arrays of STNO can be used to create directional spin-wave radiation similar to electromagnetic antennas. Combining STNO excitations with planar spin waves also creates interference patterns. We show that these interference patterns are static and have information on the wavelength and phase of the spin waves emitted from the STNO. We describe means of actively controlling spin-wave radiation patterns with the direct current flowing through STNO, which is useful in on-chip communication and information processing and could be a promising technique for studying short wave-length spin waves in different materials.
\end{abstract}

\maketitle


\section*{Introduction}

Controlling magnetization dynamics in ferromagnetic (FM) thin films is important to a new generation of wave-computing and on-chip communication devices working at high frequency and low power. Spin wave devices may complement digital semiconductor technologies and offer new possibilities for memory capacity and computational performance of particular importance as semiconductor devices miniaturization approaches fundamental limits. The main requirements for wave computation include localized sources and detectors of coherent waves in continuous or patterned propagating media. Nanometer scale electrical contacts to ferromagnetic thin films can source enough current density to generate a high-frequency dynamic response of the magnetic moments in FM films \cite{berger1996,Slonczewski1996}, resulting in emission of short wave-length spin waves \cite{Slonczewski2}.

Emission of spin waves from nanopoint contacts, so called spin-torque nano-oscillators (STNO), has been predicted theoretically \cite{Slonczewski2} and recently has also been demonstrated experimentally \cite{Demidov2010, MadamiNat2011}. Spin wave radiation from a single STNO may not be symmetric but may instead be directional due to spin wave band structures or due to dipolar fields, external fields, anisotropy film fields, or Oersted fields generated by the current in the contact \cite{demokritov_book}. For example, magnetic properties of thin films can be tailored to create preferred propagating directions and frequency band gaps, for example as done in magnonic crystals \cite{mamica_prb2011, Sklenar_apl2012}. Polycrystalline perpendicularly magnetized films may have, however, symmetric radiation patterns (in the absence of Oersted fields). In addition, STNO can be frequency and phase locked to external oscillatory signals or to other STNO frequencies under certain conditions \cite{Rippard_prl_2005, Kaka2005, mancoff_nature_2005, Pufall_prl_2006, bonin_synch, zhou, Rezende, slavin_prb2006, slavin_review}.

While applications in computation and information storage usually require a unique reference signal (a clock) that times the system and synchronizes it, wave computation offers asynchronous or poly-synchronous operation \cite{Polichronization,PWC,iopMacia}. STNO, like other wave sources, can encode information in a carrier signal by modulating combinations of amplitude, frequency, and phase. Nonlinear effects of amplitude and frequency modulation in STNO have been studied both experimentally and theoretically \cite{Pufall_apl_2005,consolo, ye, slavin_review,muduli}.  However, little work as been done thus far on phase modulation.

In this paper we investigate spin-wave patterns created by STNO and their interactions with background oscillations in perpendicularly magnetized films. We describe STNO as spin-wave antennas, derive expressions for the radiation diagrams, and discuss how to actively control the radiation patterns. We also discuss how to encode information in spin-wave radiation.

\section*{Spin-Torque Nano Oscillators}

\begin{figure}[htb]
\includegraphics[width=\columnwidth]{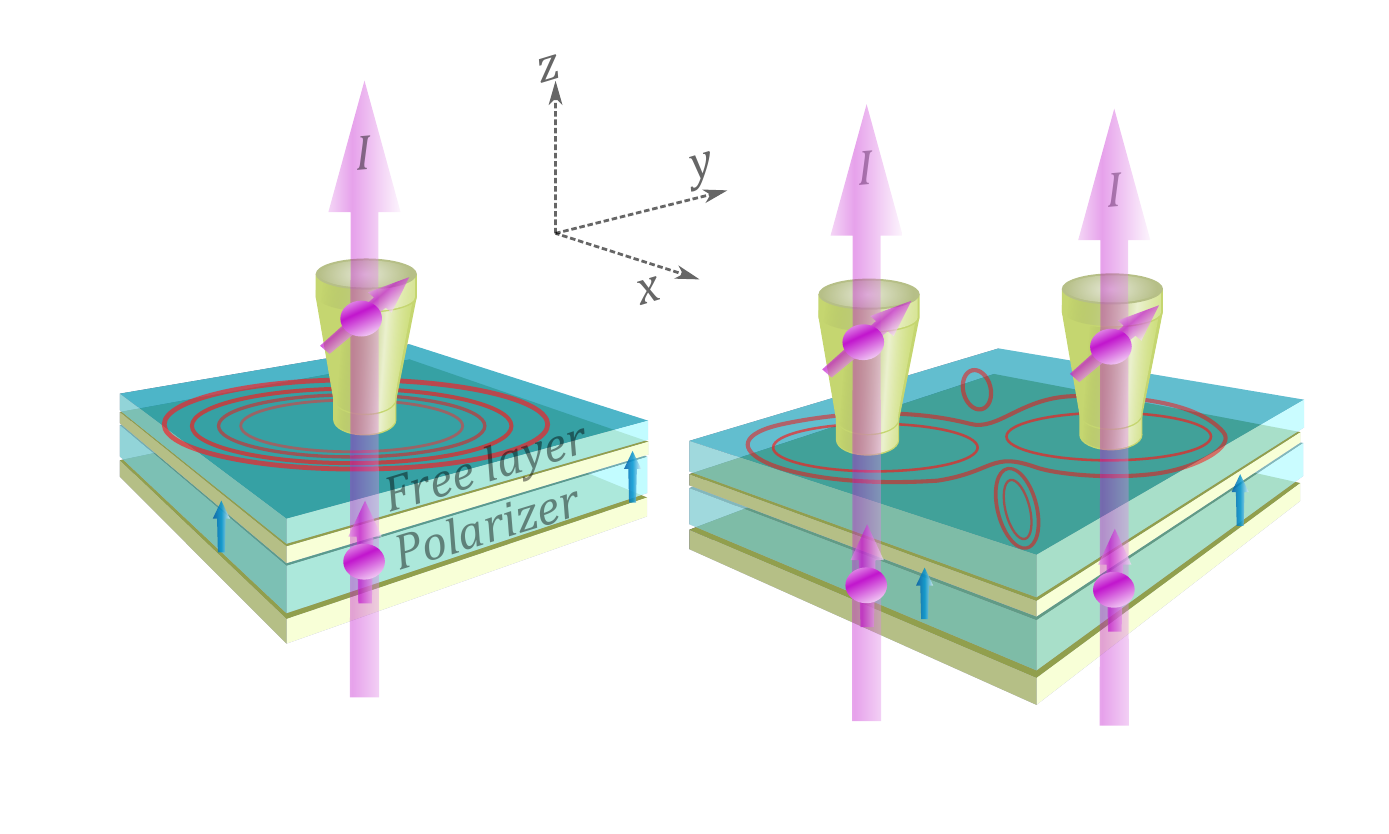}
\caption{\small{Schematic of STNO sources. An electrical charge flows through a point contact to a thin ferromagnetic layer called the free layer. The magnetic moments in the contact area precess, thereby exciting spin-waves that may propagate in the film. A double STNO arrangement (right hand side panel) creates an anisotropic spin-wave excitation that is the result of interference from two single contact excitations.}}
\label{fig1}
\end{figure}

Spin-polarized currents flowing throughout a magnetic thin film exert a torque on the background magnetization called a spin-transfer-torque \cite{Slonczewski1996, berger1996, Slonczewski2, Tsoi, Kiselev}. These polarized electrical currents encounter resistance when crossing a magnetic material that depends on the relative orientation between the current spin-polarization and the film magnetization; this is the phenomenon of
giant magnetoresistance (GMR) \cite{fertGMR,Grunberg}. STNO are nanoscale electrical contacts to a ferromagnetic thin film in a multilayer structure. Typically, dc currents greater than a critical value flow first across a thicker magnetic layer that spin polarizes them, and then pass through a thinner film where they excite magnetization dynamics. The two magnetic layers are separated by a non-magnetic layer to magnetically decouple them (see, Fig.\ \ref{fig1}).
The electrical dc-current generates a high-frequency dynamic response (1 GHz - 1 THz) in the free ferromagnetic layer and results in the emission of spin waves \cite{Slonczewski2}. STNO radiate only in two dimensions; FM thin films are considered two dimensional structures in terms of spin-wave propagation since the spin-wave wavelength and the ferromagnetic exchange length are usually much larger than the film thickness.

\section*{Mathematical Analysis}

Any perturbation of the magnetic moment initially causes the magnetic moment to precess around the direction of the effective field, $\mb{H}_{\text{eff}}$, and eventually an alignment with the effective field as a result of damping. A polarized current is modeled with an additional term in the magnetization dynamics equation \cite{Slonczewski1996,berger1996}, and when the current density exceeds a certain threshold, ($j>j_{\mbox{c}}$), a magnetic excitation develops \cite{Slonczewski1996,berger1996, Slonczewski2,Tsoi,Kiselev}.
This is described by the Landau-Lifshitz-Gilbert-Slonczewski (LLGS) equation
\beq
\begin{array}{rcl}
\ds\frac{\p \mb{M}}{\p t}&=&\ds-|\gamma| \mu_0 \mb{M} \times \mb{H_{eff}}-\alpha\frac{|\gamma|\mu_0}{M_s}\mb{M} \times (\mb{M} \times \mb{H_{eff}})
\\\\
&&\ds+\, \beta(\mb{r}) (\mb{M}\times \mb{M} \times \mb{m_p}),
\label{lle}
\end{array}
\eeq
where the precession (first term on the right-hand side) and damping (second term) are controlled by the effective field, $\mb{H_{eff}}$, which is the sum of the external magnetic field, demagnetizing field, and exchange field:
\beq
\mb{H_{eff}}=H_0\mb{z}-M_z\mb{z}+\frac{D}{|\gamma|\mu_0 M_s \hbar}\nabla^2\mb{M},
\label{Heff}
\eeq
where $M_s$ is the saturation magnetization.
Note that the exchange field magnetically couples different film locations. For example, if some area in a FM film is being excited with a polarized current from a point contact, the magnetization dynamics in the area will couple to the rest of the film and spin waves will develop in the contact neighborhood.

The spin-torque (third term in Eq.\ \ref{lle}) is controlled by the spin polarization direction of the applied current, $\mb{m}_p$. The function $\beta(\mb{r})$ is a Heaviside function defining the sizes and locations of the point contacts. $\beta(\mb{r})$ also depends on the current intensity, the layer thickness and the spin polarization \cite{iopMacia}. We consider steady state conditions, situations in which the current has been turned on for a certain time. The free magnetic layer where spin dynamics are excited is considered thin compared to the magnetic exchange length, $\lambda_{\text{ex}}$,. Therefore, we disregard variations in the magnetic moment across the film thickness ($z$ direction).

We consider a case where the free layer is perpendicularly magnetized. Since the magnetization vector $\mb{m}=\ds\frac{\mb{M}}{M_s}$ lies on the unit sphere, we focus on the lateral behavior in the free layer. We write the components of $\mb{m}$ as
$$
\mb{m}=(m,m_z)\ \ \ m=m_x+i m_y, \ \  m_z=\sqrt{1-|m|^2}.
$$
The out-of-the-plane component, $m_z$, or the absolute value of the in-plane magnetization component, $|m|^2=1-m_z^2$, describes the amplitude of the excitation and is a non-oscillating quantity. The in-plane components, $m_x$ and $m_y$, contain the frequency and phase information of the excitations.

The resulting patterns created by an STNO (or from arrays of STNO) are mostly controlled by the diffusion term in the effective field (Eq.\ \ref{Heff}). The equation for the magnetization amplitude $m$ (from Eq.\ \ref{lle} and after normalization) is a non-linear Schr\"{o}dinger equation \cite{hoefer_prl2005,iopMacia},
\beq
i\frac{\p m}{\p t}= (1+i\alpha)\nabla^2 m-f(|m|^2)m+ig(|m|^2) m,
\label{schr}
\eeq
where $f$ and $g$ are non-linear functions of the amplitude, $|m|^2$ (expressions for $f$ and $g$ are given in \cite{iopMacia})
\\
\\
A linear approximation of Eq. \ref{lle} has the form
\beq
i\frac{\p m}{\p t}= (1+i\alpha)\nabla^2 m+\omega_i\ m-i\alpha\omega_i m+i\beta(\mb{r}) m,
\label{linear}
\eeq
where $\omega_i$ is the internal frequency of the oscillator. This equation is a good starting point for studying interference patterns from different point sources. However, phase locking and stabilization of patterns in the presence of noise require analysis of Eq.\ \ref{schr}.
\\
\\
For a single nanocontact we consider a solution of the form $m(\rho,t)=\phi(\rho)e^{i\omega t}$, where $\rho$ is the radial component,
\beq
(1+i\alpha)(\p_{\rho\rho}\phi + \frac{1}{\rho} \p_{\rho}\phi)+(\omega-\omega_i) \phi + i(\beta(\mb{\rho})-\alpha\omega)\phi=0.
\label{szl}
\eeq
Slonczewski determined a linear solution using Bessel functions and appropriate boundary conditions at the nanocontact boundary \cite{Slonczewski2}.
\\
\\
\section*{Material parameters}

We now discuss typical material parameters in STNO. We have performed our calculations having in mind transition metal thin films. For a typical ferromagnetic thin film we list permalloy's most relevant parameters: $M_s\approx 860$ kA/m, $\lambda_{\text{ex}}\approx5.3$ nm, $f\approx1-30$ GHz for $H<1$ T, and the damping constant, $\alpha\approx0.01$. Spin waves are expected to propagate tens of microns in such films.

Other candidates for short wave length spin wave radiation patterns are magnetic semiconductors, a semiconductor host is magnetically doped by transition metal impurities (e.g., Ga$_{1-x}$Mn$_x$As with $x < 1-10\%$) \cite{Allan_natmat2005} or Heusler alloys, ferromagnetic metal alloys based on a Heusler phase (e.g., Cu$_2$MnAl) \cite{Picozzi}. Magnetic semiconductors and Heusler alloys have different magnetization saturation values but resonant frequencies are of the same order ($\sim$GHz); damping parameters might vary depending on the compound or the fabrication method (some Heusler alloys have been reported to have damping constants one order of magnitude smaller than transition metals). 


\section*{Radiating STNO Antennas}
\label{interfere}
The spin waves radiating from a single STNO out of the point contact (in a linear approximation) is given in terms of a Hankel function, which is a linear combination of Bessel function of the first and second kind
\beq
m(\rho,t)=H_0^2(\rho)e^{i\omega t}
\eeq
where the radial component, $\rho$, has been normalized with the wavelength, $\lambda$, which is defined by the size of the contact, $r$, being 5$r$ the first set of solutions of Eq.\ \ref{szl} \cite{hoefer_prl2005}. For large $\rho$ ($\rho\gg 2\pi$), the Hankel function is approximately
\beq
H_0^{(2)}(\rho)\approx \sqrt{\frac{2}{\pi \rho}}e^{i\frac{\pi}{4}}e^{-i\rho}=A\frac{e^{-i\rho}}{\rho}.
\label{asym}
\eeq
When different sources have a common oscillation frequency and very similar wavelengths, one can effectively describe the resulting activity patterns from STNO arrays with
\beq
\mb{m}(\rho,\theta,t) \approx D(\theta)H_0^2(\rho)\exp(i\omega t)
\eeq
where $D(\theta)$ is the energy distribution of the radiated waves \cite{iopMacia,Macia:mmm}. Although single devices may have a radiation pattern with some asymmetries, the resulting energy distributions from combining multiple sources may still be written as a single oscillator radiation multiplied by a spatial energy distribution, $D(\theta)$, known as a radiation diagram.
\begin{figure}[tb]
\includegraphics[width=80mm]{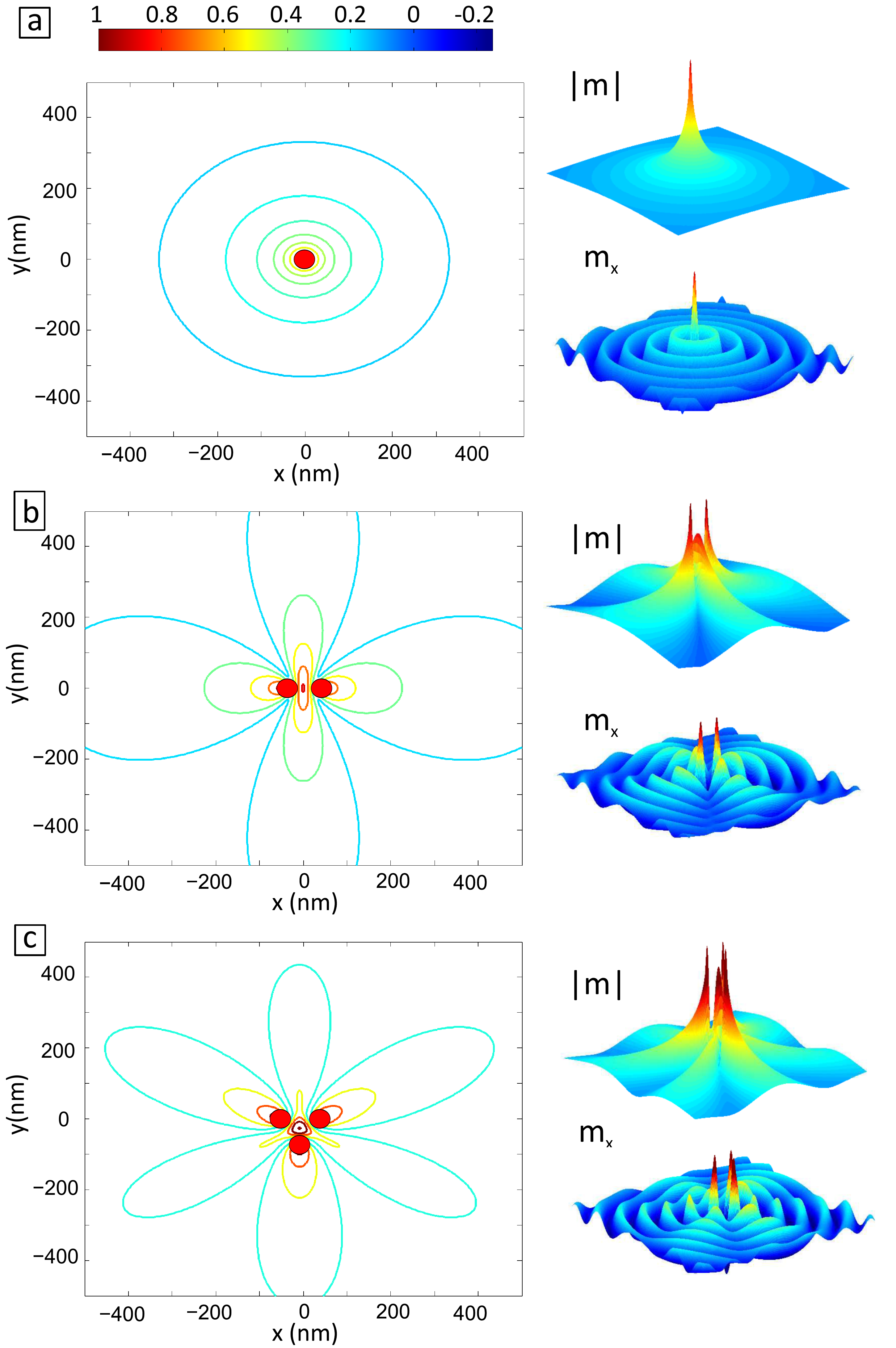}
\caption{\small{Spatial dependence of the excitation, $m$, from a single (a), a double (b), and a triple (c) STNO array. Simulation results are shown for a 500 $\times$ 500 nm$^2$ film with point contacts 40 nm in diameter, and separation between contacts of 100 nm ($\approx \lambda=2\pi/k$). The larger plots (left hand side) are contour plots of the radiated spin waves corresponding to the amplitude of the excitation, $|m|$, and the smaller 3-d graphs (right hand side) correspond to the amplitude, $|m|$, and one of the oscillating components, $m_x$.}}
\label{antennas}
\end{figure}
Figure\ \ref{antennas} shows the contour plots of the radiated spin waves for the quantity $m$ from a single (a), a double (b), and a triple (c) STNO array. For comparison we append 3-d plots of $|m|$ and $m_x$. In (a) the emission is isotropic while (b) and (c) show preferred directions of propagation.

Similar to electrical antennas, the distance between sources with respect to the wavelength, $\lambda=2\pi/k$ is crucial in controlling the radiation patterns. As required in electrical antennas, sources must radiate with exactly the same frequencies and with a set phase difference to create a steady interference pattern.

A given interference pattern created by a group of STNO may change with the relative phase, $\phi$, among oscillators. In the simplest case, two oscillators phase lock and oscillate at the same frequency having a relative phase determined by the difference in internal frequencies \cite{iopMacia}. The internal frequency of an STNO, (i.e., $\omega_1, \omega_2,$ \emph{etc}.) depends on the applied current and the local fields at the contact.

STNO can also be patterned with different shapes (ovals, rectangles, etc). However, simulations of Eq.\ \ref{lle} showed that resulting interference patterns are mostly isotropic. A physical distance between contacts (of the order of the wavelength) is required to create interference patterns.

In the next subsections we derive expressions for the radiation diagrams for simple cases of aggregated STNO; and, we calculate the radiation patterns for sets of 2 and 3 STNO and also for groups of STNO that have different oscillation phases.


\subsubsection{Radiation from 2 STNO}
\begin{figure}[tb]
\includegraphics[width=60mm]{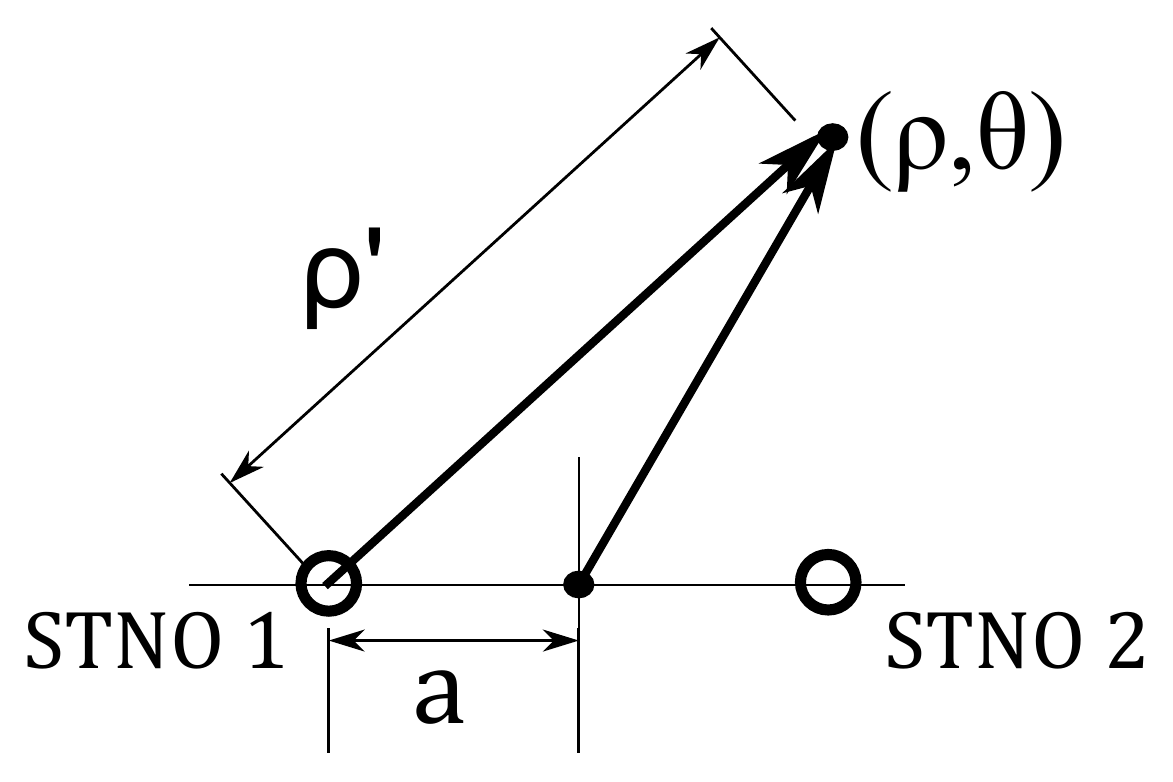}
\caption{\small{Schematic diagram of 2 STNO separated a distance $2a$. The joint coordinate system is ($\rho, \theta$) and the coordinate system for STNO 1 would be ($\rho',\theta'$)}}
\label{fig1scheme}
\end{figure}

Let us consider the simplest case for a radiation pattern from two STNO separated by a distance $2a$ and located as shown in Fig.\ \ref{fig1scheme}.
One can write the radial component, $\rho'$, in the original system of each STNO for any given point $(\rho,\theta)$ assuming the new reference system is shifted $(a,b)$ from the original one (this follows from trigonometric identities):
\beq
\begin{array}{rl}
\rho'(a,b,\theta,\rho)=&\rho-\operatorname{sgn}(a)\sqrt{a^2+b^2}\\\\
&\times\cos\left(\theta-\arctan(b/a)\right),
\end{array}
\label{rhos}
\eeq
where we have considered that $\rho\gg \sqrt{a^2+b^2}$. Following Eq.\ \ref{rhos} we obtain $\rho'=\rho-a\cos\theta$ for STNO 1 and $\rho'=\rho+a\cos\theta$ for SNTO 2.
The solution of Eq.\ \ref{linear}, $m(\rho,\theta,t)$, for the radiation is a combination of STNO 1 and STNO 2:
\beq
\begin{array}{rcl}
m(\rho,\theta,t)&=&\ds e^{i\omega t}\left(H_0^{(2)}(\rho-a\cos\theta)+H_0^{(2)}(\rho+a\cos\theta)\right)\\\\
&\ds \approx&\ds e^{i\omega t}\frac{A}{\rho}\left(e^{-i\left[\rho-a\cos\theta\right]}+e^{-i\left[\rho+a\cos\theta\right]}\right)\\\\
&=&\ds 2 e^{i\omega t}H_0^{(2)}(\rho)\cos\left(a\cos\theta\right),
\end{array}
\label{2stno}
\eeq
which is basically the solution for a single oscillator multiplied by a radiation function,
\beq
D(\theta)=2\cos(a\cos\theta),
\label{d}
\eeq
which depends only on the angle and the distance between the contacts.
\begin{figure}[tb]
\includegraphics[width=80mm]{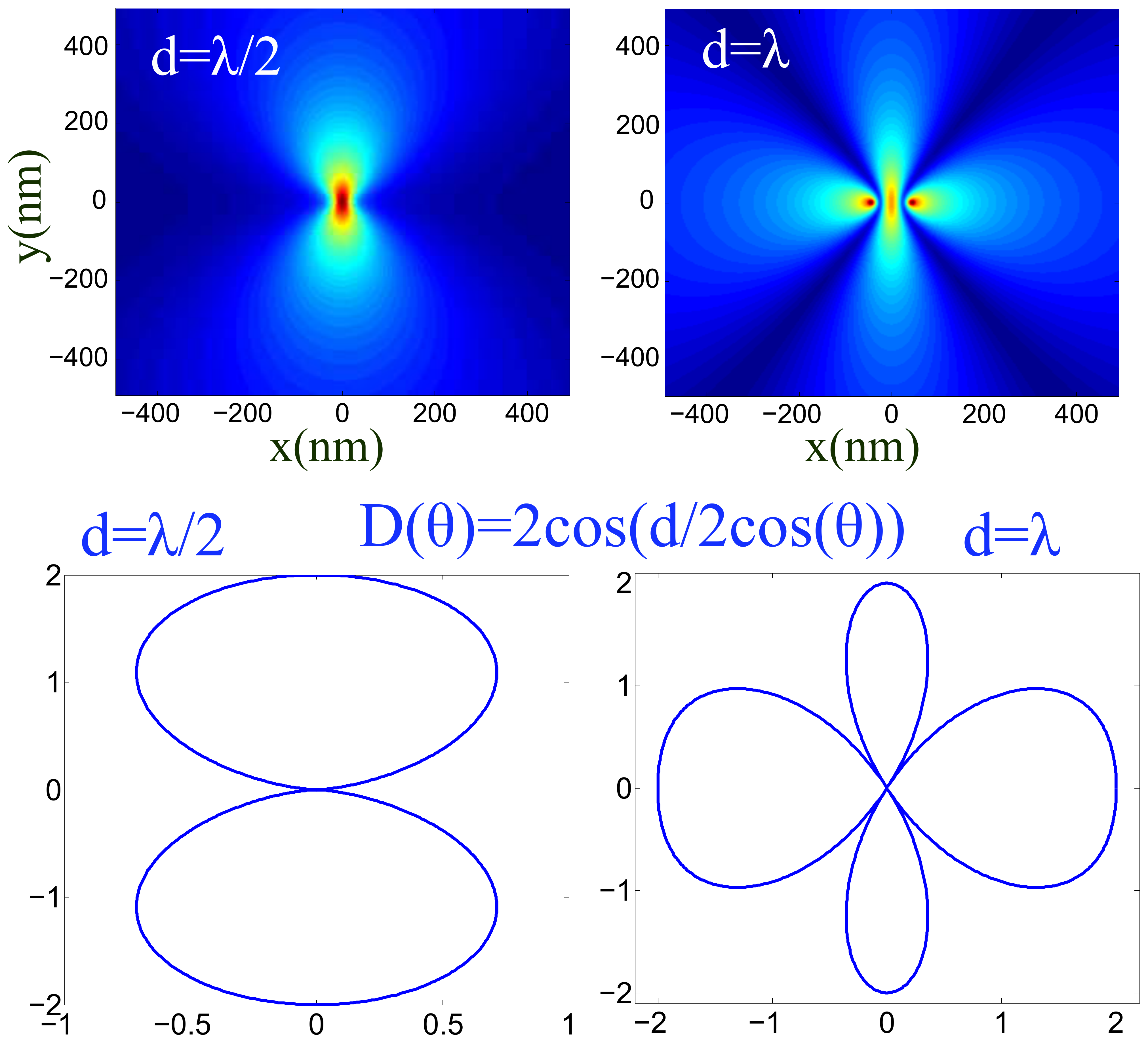}
\caption{\small{Upper panels plot $|m(\rho,\theta,t)|$ for 2 STNO as in Fig.\ \ref{fig1}; on left hand side, STNO are separated one wavelength, $\lambda$, on the right, STNO are separated half of a wavelength, $\lambda/2$. Plots correspond to a 500 $\times$ 500 nm$^2$ area with point contacts of 40 nm in diameter. Lower panels show the corresponding radiation diagram, $D(\theta)$, calculated with Eq.\ \ref{d}. Radiation diagrams are plotted in a 2-dimensional plot where $x=D(\theta)\cos(\theta)$ and $y=D(\theta)\sin(\theta)$.
}}
\label{fig:2stnos}
\end{figure}

The upper panels in Fig.\ \ref{fig:2stnos} show the spatial dependence of the excitation amplitude, $|m(\rho,\theta,t)|$, from 2 STNO positioned as shown in Fig. \ref{fig1scheme} (the amplitude $|m|$ is time independent). The lower panels show the corresponding radiation diagram, $D(\theta)$, calculated with Eq.\ \ref{d}. We have chosen two representative cases; distance, $d=2a$, between STNO to be either a half or a full wave length ($0.5\lambda$ or $\lambda$).

\begin{figure}[tb]
\includegraphics[width=80mm]{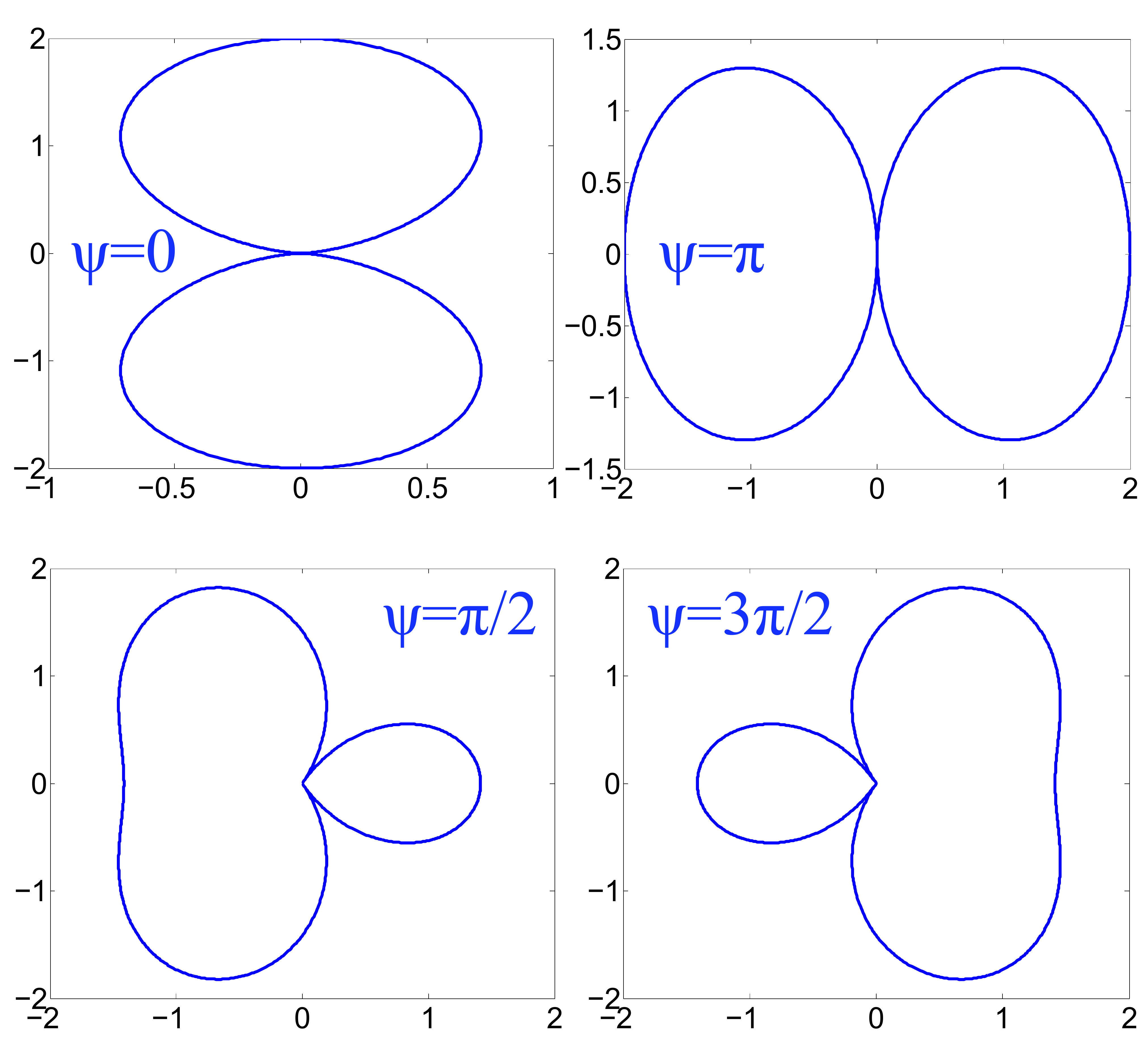}
\caption{\small{Radiation diagram, $D(\theta)$, calculated with Eq.\ \ref{phi} for $\psi=0,\pm\pi/2$ and $\pi$. STNO are separated half of a wavelength, $\lambda/2$. Radiation diagrams are plotted in a 2-dimensional plot where $x=|D(\theta)|\cos(\theta)$ and $y=|D(\theta)|\sin(\theta)$.}}
\label{fig:phi}
\end{figure}

Radiation patterns are determined by spin-wave interference from multiple sources. For two STNO, changing the phase of one of the sources, $e^{i\omega t}\rightarrow e^{i(\omega t+\psi)}$ changes the resulting radiation pattern. Radiation patterns for two STNO with a phase difference $\psi$ can be calculated and the radiation function becomes
\beq
D(\theta)=2e^{i\psi/2}\cos(a\cos\theta+\psi/2).
\label{phi}
\eeq
Figure \ref{fig:phi} shows the radiation diagrams calculated from Eq.\ \ref{phi} for different phase shifts.

\subsubsection{Radiation from 3 STNO}
\begin{figure}[tb]
\includegraphics[width=80mm]{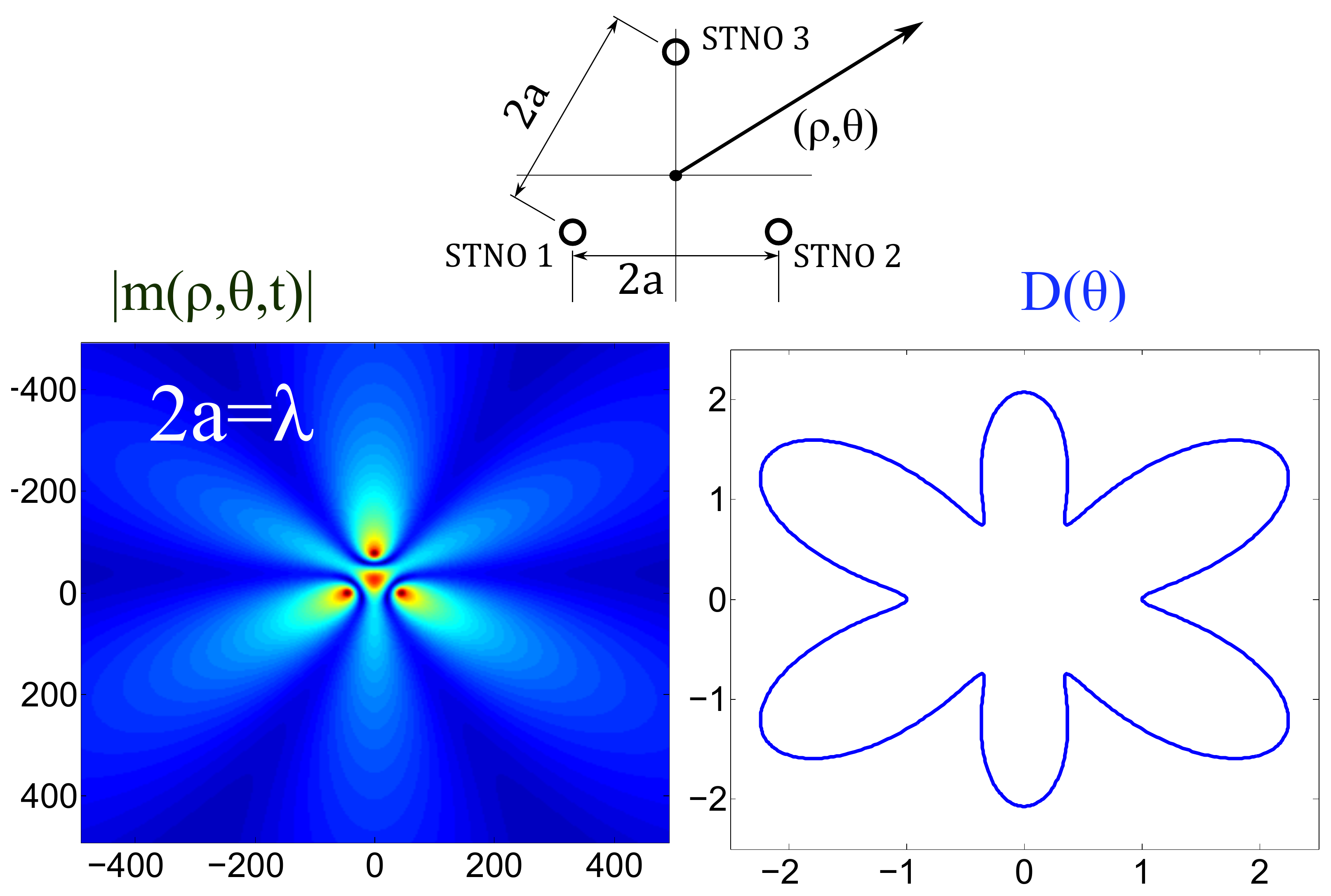}
\caption{\small{Left hand side panel shows a plot of $|m(\rho,\theta,t)|$ for 3 STNO as in Fig.\ \ref{fig:3stnos}; the plots correspond to a 500 $\times$ 500 nm$^2$ area with point contacts of 40 nm in diameter. Right hand side panel shows the corresponding radiation diagram, $D(\theta)$, calculated with Eq.\ \ref{D3stnos}. Separation between contacts is equal to the wavelength, $\lambda$.}}
\label{fig:3stnos}
\end{figure}
The radiation function for an array of 3 STNO arranged as shown in the inset of Fig.\ \ref{fig:3stnos} can be written as
\beq
\begin{array}{rcl}
D(\theta)&=&\ds e^{i\left[2a/\sqrt{3}\cos(\theta-\pi/6)\right]}+e^{-i\left[\frac{2a}{\sqrt{3}}\cos(\theta+\pi/6)\right]}\\\\
&&\ds +e^{-i\left[\frac{\sqrt{3}a}{2a}\sin\theta\right]}\\\\
&=&\ds 2\cos\left(\frac{2a}{\sqrt{3}}\cos\theta\right) e^{-i\pi/6}+e^{-i\left[\frac{\sqrt{3}}{2a}\sin\theta\right]},
\end{array}
\label{D3stnos}
\eeq
which results in a pattern that radiates in 6 directions (see, Fig.\ \ref{fig:3stnos}). Combinations with different phases allows rotation of the pattern and other complicated spin-wave excitation patterns.

\subsubsection{Radiation from n STNO}
For a general case with $n$ oscillators located at arbitrary positions one can find the expression for the radiation functions by adding the single radiation pattern from each oscillator.

\section*{Interaction with Background Oscillations}
Spin waves from STNO may also interfere with background spin-wave oscillations (e.g., with incoming planar waves). Electromagnetic radiation from electrical antennas can create planar spin-wave oscillations in FM thin films \cite{Brundle_electron1968,Bailleul_apl2003,Vlaminck_science2008}.

Here we consider the interference between a single source (STNO) and a planar wave excitation both having the same oscillating frequency. Again we take the asymptotic expression for an STNO radiation (from Eq.\ \ref{asym}), which is the same as a circular wave with wavelength $\lambda_c$,
\beq
\mb{m}_{c}(\lambda_c,\rho,t)=\ds e^{i\omega t}H_0^{(2)}\left(\frac{\rho}{\lambda_c}\right)\approx e^{i\omega t}A\frac{e^{-i\left(\frac{\rho}{\lambda_c}\right)}}{\frac{\rho}{\lambda_c}}.
\eeq
and a simple planar wave in the $x$ direction with wavelength $\lambda_p$
\beq
\mb{m}_p(\mb{\lambda_p},x,t)=e^{i\omega t}Be^{-i\left(\frac{x}{\lambda_p}\right)}.
\eeq

Let us assume there is no damping so both planar and circular excitations have constant envelopes (both in time and in space). Adding both signals we obtain
\beq
\begin{array}{rcl}
\mb{m}=\mb{m}_{c}+\mb{m}_p&=&\ds e^{i\omega t}\left(Ae^{-i\left(\frac{\rho}{\lambda_c}\right)}+Be^{-i\left(\frac{x}{\lambda_p}\right)}\right)\\\\
&=&\ds Ae^{i\omega t}e^{-i\frac{\rho}{\lambda_c}}\left(1+\frac{B}{A}e^{-i\left(\frac{x}{\lambda_p}-\frac{\rho}{\lambda_c}\right)}\right).
\end{array}
\label{planar}
\eeq

\begin{figure}[htb]
\includegraphics[width=\columnwidth]{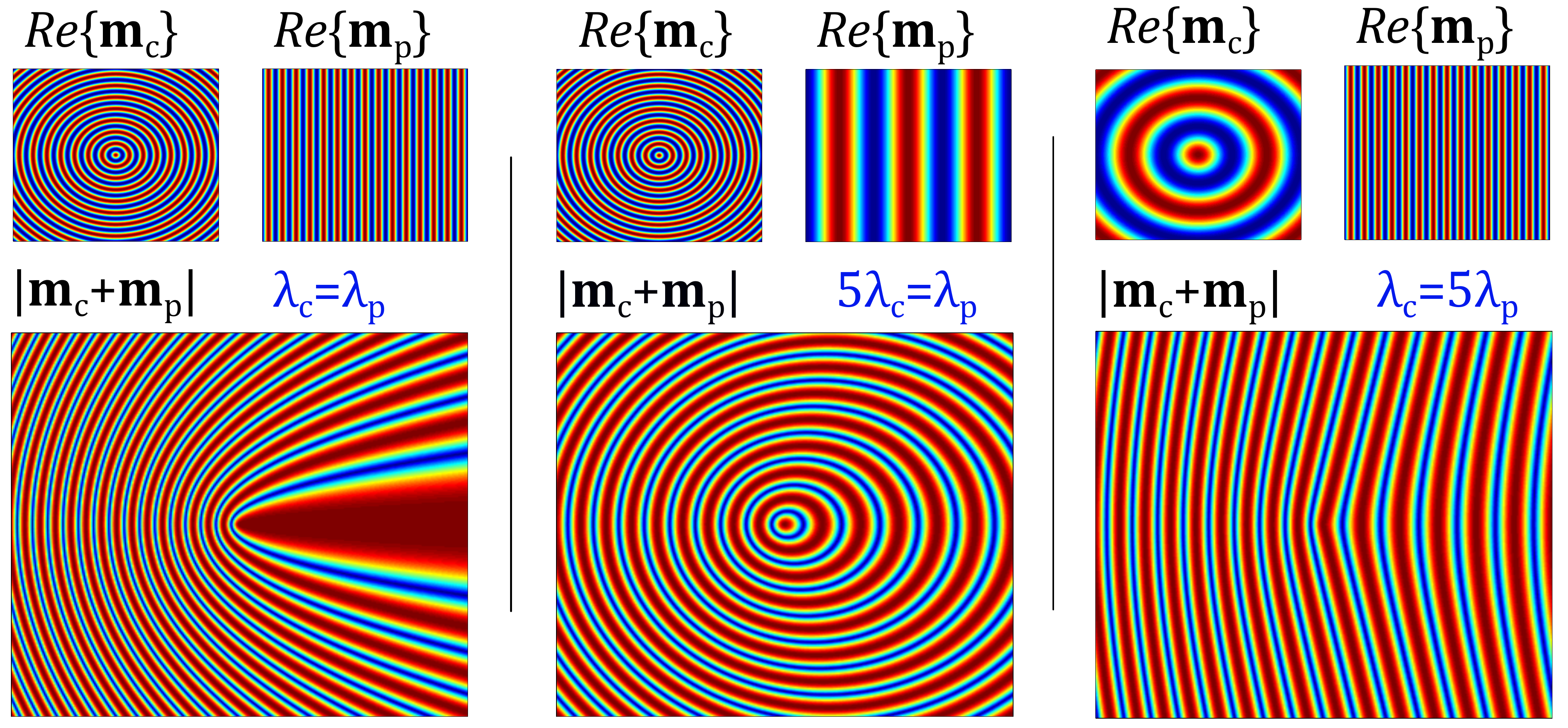}
\caption{\small{Three plots of spin-wave interference patterns between a single STNO and a planar spin wave, $\lambda_c = \lambda_p$ on the left hand side panel, $5\lambda_c = \lambda_p$ on the center panel and $\lambda_c = 5\lambda_p$ on the right hand side panel. The upper panels show the patterns of the individual excitations (time dependent patterns), the amplitude of each individual excitation is constant. Lower panels show the amplitude of the resulting interference patterns.}}
\label{figplanar}
\end{figure}

We see that adding the two waves produces a new excitation that varies in space while still being constant in time (notice that the individual excitations had constant envelopes both in time and space). We draw three different representative cases using three different wavelengths. The three cases are plotted in Fig.\ \ref{figplanar}, $\lambda_c = \lambda_p$, $5\lambda_c = \lambda_p$ and $\lambda_c = 5\lambda_p$. We first plotted the real part of both waves, which correspond to the oscillatory component $m_x$, where we identify the wavelengths, $\lambda_c$ and $\lambda_p$ respectively. The resulting interference pattern between the planar and the STNO spin wave excitations induces a pattern in the amplitude, which is a time-independent pattern that captures the structure of spin waves from both the point contact and the planar source. The dynamic response of the magnetic moment in the FM thin film is modulated by the wavelengths of the two excitations$-$a modulation of the energy.

A closer look at Eq.\ \ref{planar} shows that the interference of a planar and a circular wave creates a new wave excitation in the energy$-$or the envelope$-$ that has a wavelength that depends on the direction (taking as the origin the point source). We can write the planar excitation in the polar coordinates ($x=\rho\cos(\theta)$) centered at the point source and we obtain:
\beq
\mb{m}=\mb{m}_{c}+\mb{m}_p=e^{i\omega t}e^{-i\left(\frac{\rho}{\lambda_c}\right)}A\left(1+\frac{B}{A}e^{-i\rho\left(\frac{\cos\theta}{\lambda_p}-\frac{1}{\lambda_c}\right)}\right).
\eeq
The wavelength, $\lambda_{\text{env}}$, of the modulation envelope depends on the direction, $\theta$, and is set by
\beq
\lambda_{\text{env}}(\theta)=\frac{\lambda_p\lambda_c}{\lambda_c\cos\theta-\lambda_p},
\eeq
and the strength of the modulation depends on the relative amplitudes, $A$ and $B$, of the two waves. The smaller wavelength determines the modulation.

\section*{Detecting Spin-Wave Activity}
\label{sec:detect}

\begin{figure}[tb]
\includegraphics[width=\columnwidth]{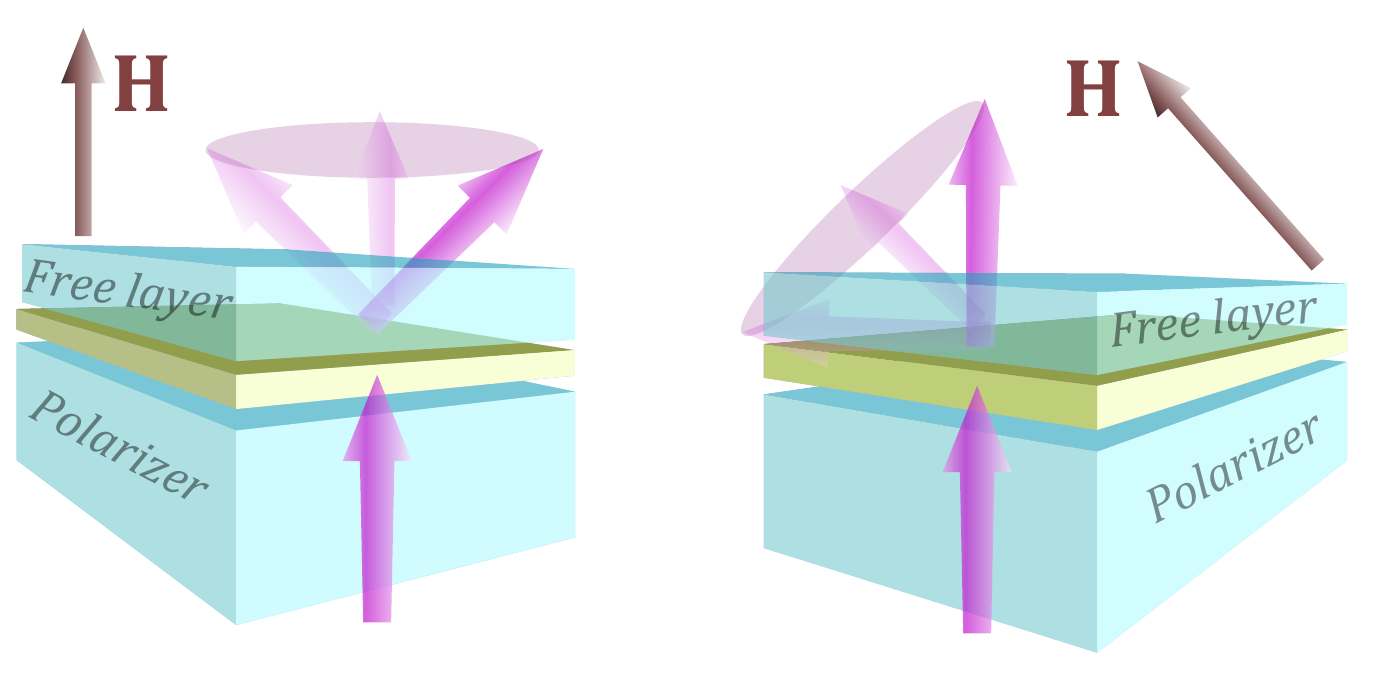}
\caption{\small{(Left hand side) Relative alignment of fixed and free layers where the resistance of the contact does not change with the oscillating magnetic moments. This configuration senses the amplitude of the wave excitation. (Right hand side) Relative alignment of a fixed and free layer where the resistance of the contact changes with the oscillating magnetic moment; this configuration extracts frequency and phase information from the wave excitations. Note that this configuration requires sufficient bandwidth in the detection circuit.}}
\label{detect}
\end{figure}

Point contacts may also be used as detectors \cite{iopMacia} of spin-wave activity in a thin film; the respective alignment of the fixed and free layers determine the resistance of the STNO \cite{fertGMR,Grunberg}. Thus, low current densities$-$that do not excite magnetization dynamics$-$through the point contacts will serve to read the state of the free layer. The magnetization of the two layers can be set in different geometries for different purposes; a geometry where the resistance of the contact is the same for all points of the oscillation trajectory (see, Fig.\ \ref{detect} left hand side panel) may serve to sense the energy or the amplitude of the spin wave excitation, $|m|$, (e.g., detecting amplitude modulation); a geometry where the resistance of the junction varies along the oscillation trajectory (see, Fig.\ \ref{detect}b) may serve to sense the frequency of the spin wave excitation (e.g., detecting frequency or phase modulation).

The giant magnetoresistance effect is weak when the orientation of one layer tilts only a few degrees, particularly when the layer magnetizations are initially collinear. Adding a magnetic tunnel junction either between the free and polarizing layer or the free layer and a separate magnetic electrode would would enhance the magnetoresistance effect \cite{houssemedine,iopMacia}.

\section*{Encoding Information in the Modulation of STNO spin waves}

STNO are spin-wave sources and detectors; their, radiating oscillatory signals may be modulated. Frequency modulation usually offers an ideal method to reduce noise in communications. However, the frequency and the amplitude of the oscillatory signals in STNO depend on the driving current; amplitude and frequency are nonlinearly connected. Experiments have shown the effect of nonlinear frequency modulation in single \cite{Pufall_apl_2005,muduli} and double STNO \cite{ye} and the effect has been described theoretically in \cite{consolo,slavin_review}.

Pulse amplitude modulation provides a case where frequency and amplitude are uncoupled; turning the spin-wave sources on and off causes a spin-wave diffusive front to propagate away from the contacts every time the STNO are pulsed. In \cite{iopMacia} it was shown that pulse modulation of an STNO source has a dissipative nature because of the spin wave diffusion (see, Eq.\ \ref{lle}). However, information can still be processed and transmitted using a simple amplitude modulation system \cite{iopMacia}.

Phase modulation may also be possible in STNO. Whenever a STNO phase locks to another STNO or to an external oscillatory signal, there is an interval of nominal frequencies where the STNO still synchronizes to the reference signal \cite{Rippard_prl_2005,Kaka2005,mancoff_nature_2005,Pufall_prl_2006}. Within this frequency interval oscillators remain synchronized but their relative phase to the reference signal changes.
In most of the spectroscopic techniques one measures the phase difference between an external signal$-$used to excite an intrinsic resonance$-$and the intrinsic resonance (e.g., Ferromagnetic resonance). One may combine arrays of synchronized STNO and tune independently their relative phases with respect of an external signal. Phase modulation may provide an interesting option for encoding information as its detection is usually simple with the help of a reference signal.

\section*{Discussion and Conclusion}


We have studied spin-wave interference patterns and directional spin-wave radiation in ferromagnetic thin film with STNO. The spatial dependence of the excitation energy in a single STNO might not be symmetric but lacks information about the wavelength of the spin-wave excitation. The spatial energy of a planar wave created by an rf-antenna also misses information about its wave length. The creation of interference patterns captures the information of wavelength from single sources. Arrays of STNO or STNO in combination with planar waves, allow energy directionality and control of the spatial energy distribution. Relative distances and positions of the STNO with respect of their wavelength produce different radiation diagrams. Patterns can also be controlled by setting the relative phase between STNO.

Communications and computation are open fields that would welcome using gigahertz excitations in solid state materials. However, a more precise control of these excitations would be needed to achieve the most basic operations. Magnonic crystals use control of the propagating media; here we explored control of radiating spin waves from STNO through interference patterns. Additionally, the studied cases of arrays of STNO and an STNO with a planar waves could serve to block certain wavelengths or to enhance others in different locations (i.e., to create bandgaps).

Another interesting outcome of this study on static interference patterns and radiation diagrams is the potential use in sensing short wavelengths of high frequency spin-wave excitations. STNO in a ferromagnet such as permalloy have excitation frequencies of tens of gigahertz and wavelengths of hundreds of nanometers; its temporal and spatial resolution are challenging and involve synchronization between the measuring equipment and the internal frequency of STNO at the nanosecond scale and space resolution at the nanometer scale. Spin-wave interference patterns from STNO arrays (or from STNO combined with planar waves) create steady patterns of the excitation amplitude (energy) that depend on the wavelength of the single sources \cite{iopMacia,Macia:mmm} and its relative phase; in such a case, a static detecting method could probe the wavelength of the excitations.


\section*{Acknowledgements}
F.M. acknowledges support from a Marie Curie IOF 253214. FCH was supported in part by Neurocirc LLC. Supported in part by ARO-MURI, Grant No. W911NF-08-1-0317.

\bibliographystyle{unsrt}
\bibliography{ms_als}


\end{document}